\documentclass[11pt]{article}
\usepackage[utf8]{inputenc}
\usepackage[T1]{fontenc}
\usepackage{lmodern}
\usepackage{geometry}
\usepackage{authblk}
\usepackage{graphicx}
\usepackage{amsmath,amssymb}
\usepackage{booktabs}
\usepackage{xcolor}
\usepackage{natbib}
\usepackage[hidelinks,colorlinks=true,linkcolor=blue]{hyperref}
\hypersetup{citecolor=blue, urlcolor=blue}

\title{pycopm: An open-source tool to tailor OPM Flow geological models}

\author[1]{David Landa-Marb\'an\thanks{Corresponding author. ORCID: \href{https://orcid.org/0000-0002-3343-1005}{0000-0002-3343-1005}}}
\affil[1]{NORCE Research AS, Bergen, Norway}

\date{28 January 2026}

\begin{document}
\maketitle

\begin{abstract}
\noindent Reservoir simulations help the energy industry make better decisions by predicting how fluids like oil, gas, water, hydrogen, and carbon dioxide will flow underground. To keep these predictions accurate, engineers often need to update geological models quickly as new information becomes available. \texttt{pycopm} is a tool designed to make this process faster and easier. It allows users to adjust geological models in several ways, such as simplifying complex grids, focusing on specific parts of a reservoir, or changing the shape and position of the model. These capabilities help engineers test different scenarios efficiently. Although \texttt{pycopm} was first used on two well-known public datasets, it has since become useful in many other situations because of its easy-to-use features and recent extensions. Today, it supports studies involving model refinement, comparing coarse and detailed models, analyzing interactions between nearby sites, and speeding up troubleshooting in large simulations. The graphical abstract is shown in Fig. \ref{fig:graphical}.
\end{abstract}

\begin{figure}[!h]
  \centering
  \includegraphics[width=\linewidth]{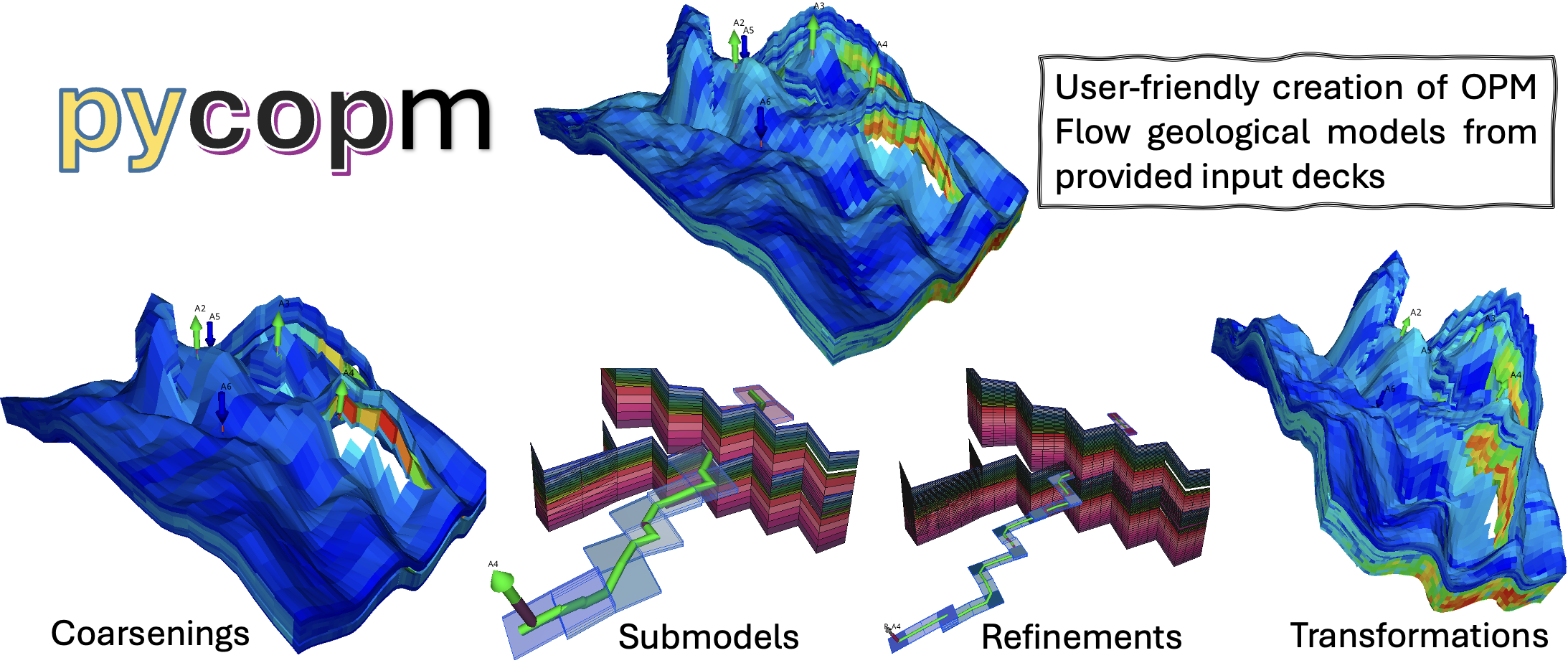}
  \caption{Graphical representation of \texttt{pycopm}'s functionality. See the \href{https://cssr-tools.github.io/pycopm/examples.html\#graphical-abstract}{online documentation} for reproduction details.}
  \label{fig:graphical}
\end{figure}

\section{Statement of need}
The first step in reservoir simulations is to \emph{choose} a simulation model, which serves as the computational representation of a geological model, incorporating properties such as heterogeneity, physics, fluid properties, boundary conditions, and wells. Once the spatial model is designed, it is discretized into cells containing average properties of the continuous reservoir model. All this information is then communicated to the simulator, which internally solves conservation equations (mass, momentum, and energy) and constitutive equations (e.g., saturation functions, well models) to perform the predictions. OPM Flow is an open-source simulator for subsurface applications such as hydrocarbon recovery, CO$_2$ storage, and H$_2$ storage~\citep{Rassmussen:2021}. The input files of OPM Flow follow the standard-industry Eclipse format. The different functionality is defined using keywords in a main input deck with extension \texttt{.DATA} and additional information is usually added in \texttt{.INC} files such as tables (saturation functions, PVT) and the grid discretization. We refer to the \href{https://opm-project.org/?page_id=955}{OPM Flow manual} for an introduction to OPM Flow and all supported keywords.

Simulation models can be substantial, typically encompassing millions of cells, and can be quite complex due to the number of wells and faults, defined by cell indices in the x, y, and z direction (i, j, k nomenclature). While manually modifying small input decks is feasible, it becomes impractical for large models. In addition, these models commonly rely on corner-point grids defined through pillars and horizons, and they may include further geometric modifications specified through deck keywords. Such representations are not intuitive to manipulate, particularly for users who are not familiar with the internal structure of simulation decks.

These challenges inspired the development of \texttt{pycopm}, a user-friendly Python tool designed to tailor geological models from provided input decks. \texttt{pycopm} is intended for researchers, engineers, and students who need to apply model transformations such as coarsening, refinement, submodel extraction, and structural adjustments. The coarsening and refinement capabilities are especially relevant in current workflows, since multi-fidelity modeling has become an active and widely adopted research direction that requires the flexibility to generate models with different levels of complexity.

\section{State of the field}
Two key properties in a reservoir model are its storage capacity, measured by pore volume, and the ability of fluids to flow between cells, known as transmissibilities. Therefore, these properties must be properly handled when generating a new model. While grid refinements and transformations do not pose a significant issue, submodels and grid coarsening present challenges due to lack of unique methods for addressing these properties. In other words, the approach depends on the specific model, and while there are a few methods in the literature, this remains an active area of research.

\href{https://github.com/OPM/opm-upscaling}{opm-upscaling} is part of the \href{https://opm-project.org/?page_id=23}{OPM initiative}, and provides a set of C++ tools focused on single-phase and steady-state upscaling of capillary pressure and relative permeability. However, it does not include functionality for grid refinement or affine transformations, and its upscaling routines operate mainly on the grid structure. As a result, users must manually adjust the remaining components of the deck to match the new i, j, and k indices. Examples include updating the locations of wells, numerical aquifers, and other model elements.

\href{https://resinsight.org}{ResInsight} is an open-source C++ tool designed for postprocessing reservoir models and simulations. It can export modified simulation grids and supports operations such as grid refinement and submodel extraction. Nevertheless, it does not generate an updated input deck reflecting the modified i, j, and k coordinates. It also lacks built-in capabilities for model coarsening or for applying general affine transformations. Users on macOS may encounter installation challenges, and the software has difficulty handling models with very small cell dimensions (below 1 mm) or with very large cell counts (greater than 100 million).

To the author's knowledge, prior to the development of \texttt{pycopm} there was no integrated Python-based solution that combined coarsening, refinement, submodel extraction, and geometric transformations for modifying geological models compatible with OPM Flow. Python offers a significantly more accessible environment than C++, which lowers the entry barrier for researchers, engineers, and students who need flexible model manipulation tools. \texttt{pycopm} offers flexibility in selecting different approaches, allowing end-users to compare methods and choose the one that best fits their needs. For additional information about the different approaches implemented in \texttt{pycopm} for coarsenings, refinements, submodels, and transformations, see the \href{https://cssr-tools.github.io/pycopm/theory.html}{theory} in \texttt{pycopm}'s documentation.

\section{Software design}
\texttt{pycopm} leverages well-established and excellent Python libraries. The Python package NumPy~\citep{2020NumPy-Array} forms the basis for performing array operations. The pandas package~\citep{the_pandas_development_team} is used for handling cell clusters, specifically employing the methods in \texttt{pandas.Series.groupby}. The Shapely package~\citep{Gillies_Shapely_2025}, particularly the \texttt{contains\_xy}, is fundamental for submodel implementation to locate grid cells within a given polygon. To parse the output binary files of OPM Flow, the \href{https://pypi.org/project/opm/}{opm} Python libraries are utilized. The primary methods developed in \texttt{pycopm} include handling of corner-point grids, upscaling transmissibilities in complex models with faults (non-neighbouring connections) and inactive cells, projecting pore volumes on submodel boundaries, interpolating to extend definition of i, j, k dependent properties (e.g., wells, faults) in grid refinement, and parsing and writing input decks.

Interaction with the tool is performed through a terminal executable named \texttt{pycopm}, which provides a set of command-line flags (27 at the time of writing; see the online documentation for the \href{https://cssr-tools.github.io/pycopm/introduction.html#overview}{current list}). These flags control the desired functionality, such as specifying the input deck, defining how the model should be modified, and selecting the output file name. This design enables users to chain multiple operations by further editing the generated decks. For example, a user may refine a model first and subsequently extract a submodel. An illustrative example is provided in \href{https://github.com/cssr-tools/pycopm/blob/main/tests/test_4_submodel.py}{test\_4\_submodel.py} in the project repository. Advanced users who are familiar with Python can access the underlying functionality directly through Python scripts. This provides greater flexibility for integrating the tool into more sophisticated workflows and for customizing model transformations to meet specific research or engineering needs.

\section{Research impact statement}
\texttt{pycopm} is already being adopted across several projects at NORCE Research AS and Equinor ASA. The user base is growing, with increased activity observed across multiple locations. GitHub traffic insights also show a steady rise in repository clones, indicating expanding interest and usage.

The software has supported several research publications, including:
\begin{itemize}
  \item \cite{2024sandve}, where it was used to coarsen the Drogon model for history matching,
  \item \cite{nilsen2025}, which applied coarsening of the same model to optimize operations using wind energy,
  \item \cite{Sandve2025}, where submodel extraction and coarsening were applied to the Troll aquifer model to analyze pressure interference, and
  \item \cite{landamarban2025}, which used coarsening of the Troll aquifer model to optimize well placement in CO$_2$ storage simulations.
\end{itemize}

\noindent \texttt{pycopm} is part of the software suite developed within the \href{https://cssr.no}{Centre for Sustainable Subsurface Resources} and maintained under the \href{https://github.com/cssr-tools}{cssr-tools} GitHub organization. A key objective of these tools is to support research outputs that adhere to the FAIR principles (Findable, Accessible, Interoperable, Reusable) originally formalized in \cite{Wilkinson2016}. These principles have not been consistently implemented in subsurface research in recent years~\citep{liu2025}, limiting the long-term impact and reproducibility of published results. To address this, significant effort has been dedicated to building comprehensive online documentation that enables users to reproduce figures, tables, and computational workflows from recent publications. For example, the \href{https://cssr-tools.github.io/expreccs/tccs-13.html#}{TCCS-13} documentation includes step-by-step terminal commands required to generate the results presented in \citep{landamarban2025}. This ensures that published work is not only transparent but also directly reusable by other researchers, enhancing scientific rigor and accelerating future developments.

Looking ahead to increase the research impact, the plan for \texttt{pycopm}'s future development includes extending its functionality to support additional keywords from input decks beyond those in geological models, which \texttt{pycopm} has been successfully tested on (\href{https://github.com/OPM/opm-tests/tree/master/drogon}{Drogon}, \href{https://github.com/OPM/opm-tests/tree/master/norne}{Norne}, \href{https://co2datashare.org/dataset/smeaheia-dataset}{Smeaheia}, \href{https://github.com/OPM/opm-data/tree/master/spe10model2}{SPE10}, \href{https://arxiv.org/abs/2508.08670}{Troll aquifer model}). This support will be added as \texttt{pycopm} is applied in further models, and external contributions to the tool are welcomed. Additionally, extending \texttt{pycopm}'s capabilities includes implementing a feature to generate a single input deck by combining geological models from different input decks.

\section{AI usage disclosure}
No generative AI tools were used in the development of this software. Microsoft M365 Copilot (powered by a GPT-5–class large language model developed by Microsoft) was used to check and improve the writing of this manuscript.

\section*{Acknowledgements}
The author acknowledges funding from the \href{https://cssr.no}{Center for Sustainable Subsurface Resources (CSSR)}, grant nr. 331841, supported by the Research Council of Norway, research partners NORCE Norwegian Research Centre and the University of Bergen, and user partners Equinor ASA, Harbour Energy, Sumitomo Corporation, Earth Science Analytics, GCE Ocean Technology, and SLB Scandinavia. The author also acknowledges funding from the \href{https://www.norceresearch.no/en/projects/expansion-of-resources-for-co2-storage-on-the-horda-platform-expreccs}{Expansion of Resources for CO$_2$ Storage on the Horda Platform (ExpReCCS) project}, grant nr. 336294, supported by the Research Council of Norway, Equinor ASA, A/S Norske Shell, and Harbour Energy Norge AS. The author extends sincere gratitude to Tor Harald Sandve and Sarah Gasda for their valuable insights that significantly enhanced the development of \texttt{pycopm}.

\bibliography{paper}

\end{document}